# Surface Waves and Axoplasmic Pressure Waves in Action Potential Propagation: Fundamentally Different Physics or Two Sides of the Same Coin?

Marat M. Rvachev [*,‡] and Benjamin Drukarch [†,§]
[*]*Independent Researcher, New York, NY, USA*
[†]*Department of Anatomy and Neurosciences, Amsterdam UMC*
*Amsterdam Neuroscience, Amsterdam, The Netherlands*
[‡]*rvachev@alum.mit.edu*
[§]*b.drukarch@amsterdamumc.nl*

In this commentary, we argue that El Hady and Machta's "surface wave" model[1] for mechanical waves accompanying action potential (AP) propagation describes the same underlying process as the "axoplasmic pressure wave" model introduced earlier by Rvachev.[2] Both models describe mechanical modes that store potential energy in the elastic components of the axon (axonal membrane, cytoskeleton, bulk axoplasmic deformation), with kinetic energy carried by the axoplasmic fluid and axoplasmic viscosity playing a significant role. The "surface wave" model quantitatively considers driving by the traveling electrical depolarization wave of the AP, whereas the "axoplasmic pressure wave" model qualitatively considers driving not only by the AP's electrical depolarization but also by other mechanisms, such as cytoskeletal actomyosin contractility. In addition, the "axoplasmic pressure wave" model considers mechanisms for synchronizing the depolarization wave and the pressure wave. Although derived using different approaches, the two models yield identical dependencies for the mechanical modes in key limits. The confusion in the literature, which treats these models as describing distinct processes, needs to be resolved to improve comprehensive understanding of the AP phenomenon and to guide future research.

For more than seventy years, action potentials (APs) propagating along the neuronal axon surface have been modeled and experimented on largely within the confines of the electrical framework introduced by Hodgkin and Huxley (HH) in the 1950s.[3] In recent decades, however, experimental evidence is mounting demonstrating that propagating APs are accompanied by nonelectrical changes in a number of biophysical properties of the axon, such as temperature, axonal swelling, and changes in intracellular pressure. These, (largely) reversible, nonelectrical manifestations cannot be straightforwardly accommodated within the irreversible, electrical, HH-based, framework of AP generation and propagation. This prompted renewed efforts to (mathematically) model AP propagation accounting for both electrical as well as nonelectrical signs.[4] A popular way in which this has been approached is to refine and extend the HH model through additional equations aimed at coupling the thermal, mechanical, and other nonelectric manifestations of the nerve signal with the electric ones, although more radical departures from the HH framework have also been proposed.[5]

[‡]Corresponding author.

An explicit challenge to the HH model came from Heimburg and Jackson,[6] who in 2005, building on earlier work by Kaufmann,[7] proposed that APs are adiabatic density pulses (solitons) in the axonal membrane, propagating via nonlinear, dispersive mechanics. In 2007, they extended their model, publishing in this journal,[8] to show how anesthetics affect membrane phase behavior and pulse propagation.

In 2010, also in this journal, Marat Rvachev presented an "axoplasmic pressure wave" model, suggesting that HH-type AP propagation is accompanied by a pressure pulse traveling through the axoplasm along the axon's length.[2] The Rvachev model proposed a mechanism that allows the electrical and mechanical waves to synchronize via stretch modulation of voltage-gated ion channels. For pressure pulse generation and amplification, several mechanisms were suggested, including a $Ca^{2+}$-driven process that may act via actomyosin contractility, as well as a simpler mechanism of electro-mechanical coupling of the membrane to the HH voltage spike, as further discussed below.

This was followed in 2015 by Ahmed El Hady and Benjamin Machta, who proposed a "surface wave" model that considered the mechanical (and other nonelectrical) aspects of the AP as driven by the electrical aspect, irrespective of how the electrical wave is generated.[1] In their words, "Our model does not require an underlying theory of how this electrical component arises. We emphasize that any travelling electrical wave will induce a co-propagating mechanical wave . . .."[1]

However, in their effort to promote the novelty of their model, El Hady and Machta moreover stated that "Our model differs from most other models of the mechanical response existing in the literature, in that it is electrically driven by the depolarization wave that forms the AP. Where driving has been explicitly considered, it has been taken to arise from actomyosin contractility."[1] With this latter statement, El Hady and Machta specifically refer to the Rvachev model.[2] Contrary to the claim of El Hady and Machta, however, the "axoplasmic pressure wave" model does specifically consider the electrical driving of axoplasmic pressure waves, albeit, along with actomyosin contractility. It states: "We also posit that generation and amplification of the axoplasmic pressure pulse may proceed through electro-mechanical coupling such as voltage-induced membrane movement[9] resulting from the HH voltage spike. ... any axoplasmic disturbance resulting from a propagating AP (e.g., caused by the influx of extracellular $Ca^{2+}$ ions and their presumed action on acto-myosin cytoskeletal elements, or the disturbance caused by voltage-induced membrane movement[9]), should accumulate into a larger, shock-like axoplasmic pressure wave such as we propose."[2] Therefore, in our opinion, the statement by El Hady and Machta that "Where driving has been explicitly considered, it has been taken to arise from actomyosin contractility" is factually incorrect. We believe that, in fact, the two models describe the same mode of pressure wave propagation, termed "surface waves" in one model and "axoplasmic pressure waves" in the other. Both the surface waves and the axoplasmic pressure waves are interrelated and form part of the same process in which one influences the other, similar to the ideas put forward by Wilke at the start of the twentieth century[10] and as suggested by the similarity in geometries (Fig. 1).

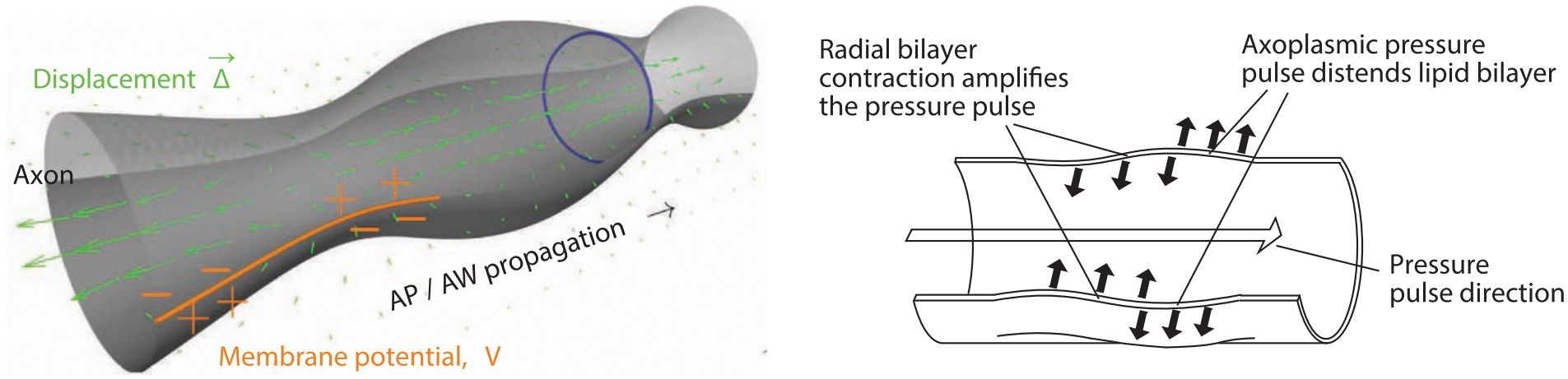

(a) (b)

Fig. 1. Geometries of the two models. Both models involve pressure waves accompanying an AP, propagating in the axoplasm and distending the membrane. (a) El Hady and Machta model. Reproduced with permission from Ref. 1. (b) Rvachev model. Although formulated as a pressure-pulse-driven membrane distention followed by contraction, the model yields the same results for other perturbation sequences, such as contraction followed by distention. Reproduced with permission from Ref. 2.

Thus, in both the high-viscosity ($\alpha \ll 1$) and low-viscosity ($\alpha \gg 1$) regimes, the two models yield the same velocity dependence for pressure pulse propagation (note that the parameter $\alpha$ in the El Hady and Machta model, "$\alpha = \rho_{3D} r_0^2 \omega/\eta$," corresponds to $\alpha^2$ in the Rvachev model, "$\alpha \equiv R\sqrt{\omega\rho/\mu}$"):

- High viscosity ($\alpha \ll 1$): The pressure pulse propagation velocity varies as the square root of the membrane expansion modulus, axon radius, and pulse frequency, and inversely as the square root of viscosity ("$\ldots c_{pr} \sim \sqrt{\kappa r_0 \omega/\eta} \ldots$" (Ref. 1, Results, 6th paragraph); "$v_{gr} = \sqrt{2R\omega K/3\mu}$" (Ref. 2, Eq. (6))).
- Low viscosity ($\alpha \gg 1$): The velocity varies as the square root of the membrane expansion modulus and inversely as the square root of the axoplasm density and axon radius ("$\ldots c_{pr} \sim \sqrt{\kappa/\rho_{3D} r_0} \ldots$" (Ref. 1, Results, 6th paragraph); "$v_0 = \sqrt{K/2\rho R}$" (Ref. 2, Eq. (1), neglecting axoplasm compressibility)).

While there are some relatively minor differences between the model formulations, they do not fundamentally alter the underlying physics of passive pressure wave propagation:

(1) El Hady and Machta specifically punish radial elastic deformations, motivated by experiments showing evenly spaced actin rings.[11]
(2) El Hady and Machta explicitly incorporate the effects of extracellular fluid.
(3) Rvachev accounts for the bulk compressibility of the axoplasm.

In addition, a key idea in the Rvachev model is the stretch modulation of voltage-gated sodium (Nav) channels, which allows the electrical and mechanical waves to co-propagate at a velocity higher than in the classical purely electrical Nav activation mechanism and may be close to the velocity of the mechanical wave, regardless of whether the mechanical driving force is actomyosin contractility, voltage-induced membrane movement, or another axoplasmic disturbance caused by the AP. The model treats all these driving forces qualitatively, while the El Hady and Machta model provides quantitative estimates (including thermal aspects) for voltage-induced membrane movement—estimates that remain valid even if the electrical and mechanical components propagate at different velocities.

In fact, El Hady and Machta state: "We emphasize that the driven waves we consider will travel at the speed of the electrical AP that drives them, $c_{AP}$, and not at their undriven propagation speed $c_{pr}$."[1] However, a mechanical pressure wave or pulse cannot be constrained to propagate at an arbitrary $c_{AP}$, just as when a loudspeaker vibrates, the resulting sound waves propagate at the speed of sound in the medium (air), not at the velocity of the loudspeaker. That the membrane displacement (and the associated axoplasmic pressure pulse) propagates ahead of the AP in cases where $c_{pr} > c_{AP}$ can, in fact, be seen in El Hady and Machta's results (Figs. 2(a)–2(f)), consistent with the Rvachev model. While the experimentally observed magnitude of the membrane displacement appears to be well described by electromechanical coupling with the AP in the El Hady and Machta model, the velocity of the driven mechanical modes is, in our opinion, also similar to the axoplasmic pressure wave modes in the Rvachev model, which, like the El Hady and Machta model, assumes that the duration of the axoplasmic pressure pulse is comparable to that of the AP (both models, however, suffer from imprecise knowledge of the mechanical properties of axons). Although the El Hady and Machta model can be considered more general than the Rvachev model, in the sense that it does not necessarily prescribe a functional role for axoplasmic pressure waves (particularly in cases where $c_{pr} < c_{AP}$), El Hady and Machta, like Rvachev, focus primarily on the potentially functional implications of cases where $c_{pr} > c_{AP}$, stating: "... it seems likely that biology would take advantage of this co-propagating information, especially if real neurons are indeed in the regime where $c_{pr} > c_{AP}$ in which the mechanical component arrives before the electrical AP. ... We foresee future theoretical and experimental work will highlight the extent to which AWs (action waves) play a functional role in neuronal information processing."[1] On the other hand, we expect that other pressure pulse-generating processes driven by the AP could produce a qualitatively similar picture to that of the Rvachev model.

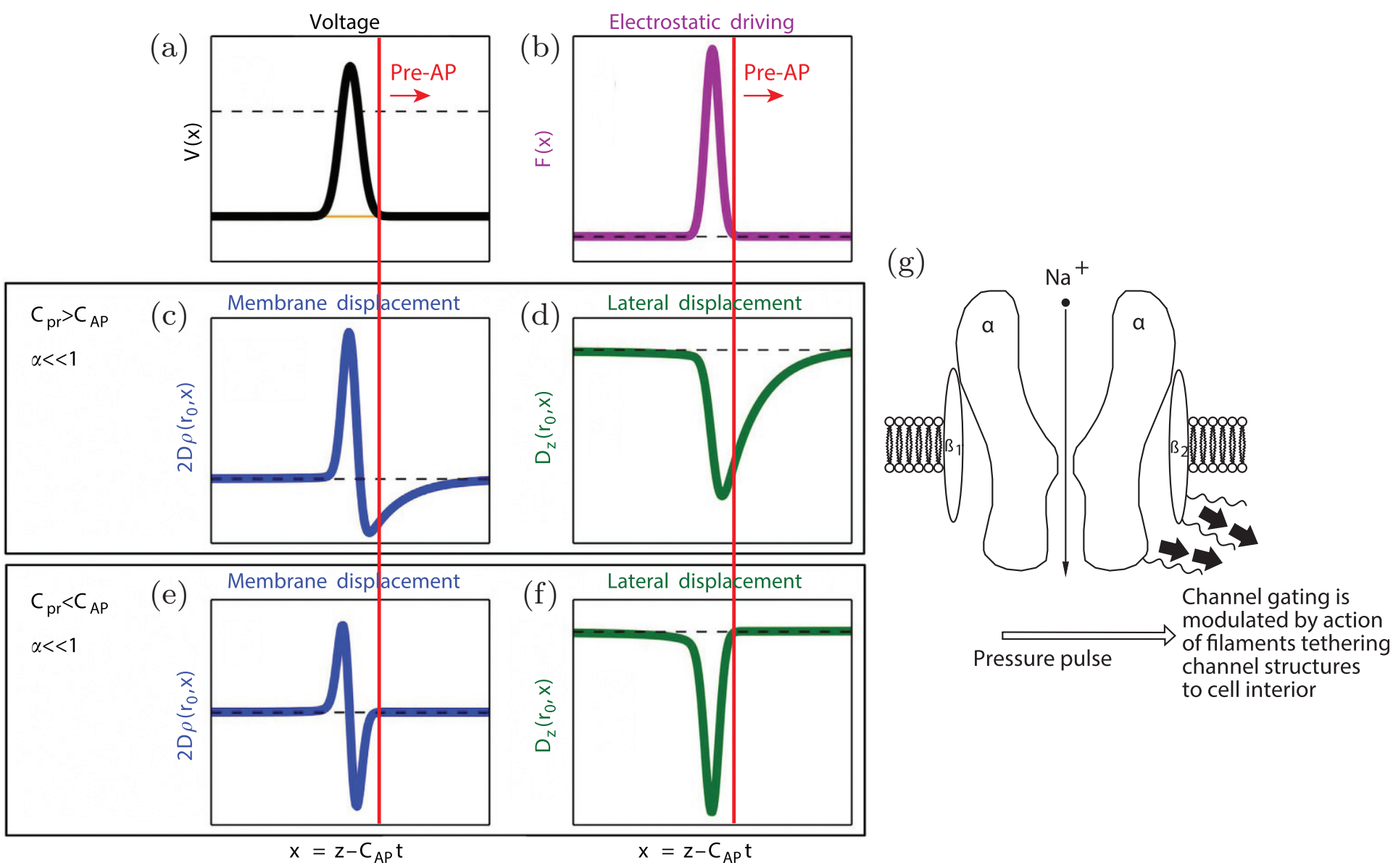

Fig. 2. (a)–(f) Displacements induced by electrical driving in the El Hady and Machta model. The assumed Gaussian wave of depolarization of the AP (a) leads to a traveling force on the mechanical modes (b). The radial displacement of the membrane (c),(e) and the average longitudinal displacement inside the axon (d),(f) are shown for parameter regimes where $c_{pr} > c_{AP}$ (c),(d) and $c_{pr} < c_{AP}$ (e),(f), in the high-viscosity regime ($\alpha \ll 1$). When $c_{pr} > c_{AP}$, the displacements, and consequently the associated axoplasmic pressure variation, precede the arrival of AP depolarization (region to the right of the vertical red lines). Modified with permission from Ref. 1. (g) Reproduced with permission from Ref. 2, where it was proposed that "… it is plausible that Nav channels are modulated by direct mechanical links (such as filaments) tethering the ion channel structures to the interior cytoskeleton that is perturbed by the propagating pressure pulse."

In addition to the above, in describing their model, El Hady and Machta, in our opinion, in effect restate several proposals from the Rvachev model without proper attribution:

- "Interestingly, most neurons are in the regime where $c_{pr} \sim r^{1/2}$ as predicted by the cable theory." (The square root dependence of pressure pulse velocity is derived in the Rvachev model in unmyelinated axons: "$v_{gr} = \sqrt{2R\omega K/3\mu}$" (Ref. 2, Eq. (6))).
- "AWs could feed back and influence the electrical component of the AP."
- "Mechanics could influence electrical properties … by influencing the gating of voltage-gated ion channels."

- "More plausibly relevant for the AP, a channel could directly sense displacements by connecting to a cytoskeletal element through a tether whose length is coupled to channel activity." (Compare with Fig. 2(g), reproduced from Ref. 2.)

Proper mathematical modeling and physical explanation of the propagating AP as a multiphysical phenomenon mediating signal transmission and information processing in the nervous system remains a contentious topic debated by various groups of theoretical and experimental neurophysicists/physiologists.[5] In these discussions, the "surface waves" model of El Hady and Machta and the earlier "axoplasmic pressure waves" model of Rvachev continue to be treated as describing distinct physical processes,[12] sometimes used together in formulating a general unifying account of AP propagation.[13] However, as argued here, this is based on a misconception introduced by El Hady and Machta, misattributing aspects of the "Rvachev model" to their own work. In light of current confusion, the many unfortunate misunderstandings which already riddle the debate on the physics of the AP,[5] and the importance of the propagating AP as a central concept in current scientific understanding of neuronal signaling, this misconception should be addressed and resolved. Doing so would also contribute to refining recent ideas about the potential role of intracellular pressure integration in dendritic signal processing.[14]

### Acknowledgments

We thank the reviewers for their insightful comments and constructive suggestions, which substantially improved the paper.

## ORCID

Marat M. Rvachev https://orcid.org/0009-0007-1380-1703
Benjamin Drukarch https://orcid.org/0000-0002-2417-8888